\documentclass[%
 reprint,
 amsmath,amssymb,
 aps,
]{revtex4-1}

\usepackage{graphicx}
\usepackage{dcolumn}
\usepackage{bm}
\usepackage{color}

\begin{document}

\preprint{APS/123-QED}

\title{Plasmonic Nanolenses Produced by Cylindrical Vector Beam Printing for Sensing Applications.}

\author{S.A. Syubaev$^{1,2}$}
\author{A.Yu. Zhizhchenko$^{1,2}$}
\author{D.V. Pavlov$^{1,2}$}
\author{S.O. Gurbatov$^{1,2}$}
\author{E.V. Pustovalov$^{1}$}
\author{A.P. Porfirev$^{3,4}$}
\author{S.N. Khonina$^{3,4}$}
\author{S.A. Kulinich$^{1,5}$}
\author{J.B.B. Rayappan$^{6}$}
\author{S.I. Kudryashov$^{7,8}$}
\author{A.A. Kuchmizhak$^{1,2}$}

\email{alex.iacp.dvo@mail.ru}

\affiliation{$^1$Far Eastern Federal University, Vladivostok, Russia}
\affiliation{$^2$Institute of Automation and Control Processes, Far Eastern Branch, Russian Academy of Sciences, Vladivostok, Russia}
\affiliation{$^3$Samara National Research University, Samara, Russia}
\affiliation{$^4$IPSI RAS - Branch of the FSRC "Crystallography and Photonics" RAS, Samara, Russia}
\affiliation{$^5$Department of Mechanical Engineering, Tokai University, Hiratsuka, Kanagawa, Japan}
\affiliation{$^6$School of Electrical and Electronics Engineering, SASTRA Deemed University, Thanjavur, Tamil Nadu, India}
\affiliation{$^7$Lebedev Physical Institute, Russian Academy of Sciences, Moscow, Russia}
\affiliation{$^8$National Research Nuclear University MEPhI, Moscow, Russia}

\date{\today}

\begin{abstract}

Here, we demonstrate a simple direct maskless laser-based approach for fabrication of back-reflector-coupled plasmonic nanorings arrays. The approach is based on delicate ablation of an upper metal film of a metal-insulator-metal (MIM) sandwich with donut-shaped laser pulses followed by argon ion-beam polishing. After being excited with a radially polarized beam, the as-prepared MIM configuration of the nanorings permitted to realize efficient nanofocusing of constructively interfering plasmonic waves excited in the gap area between the nanoring and back-reflector mirror. For optimized geometric parameters of such MIM structure, substantial enhancement of the electromagnetic near-fields at the center of the ring within a single focal spot with the size of 0.37$\lambda^{2}$ can be achieved, which is confirmed by Finite Difference Time Domain (FDTD) calculations, as well as by detection of enhanced PL signal from adsorbed organic dye molecules. The simple large-scale and cost-efficient fabrication procedure used, along with relatively good tolerance to excitation beam misalignment, make the proposed structures promising for realization of various nanophotonic and biosensing platforms that utilize cylindrical vector beam as a pump source.

\end{abstract}

\pacs{Valid PACS appear here}
\maketitle

\textbf{INTRODUCTION.}

Surface plasmons (SPs) are collective oscillations of free-electron plasma that can be excited on a metal-dielectric interface with a properly polarized radiation at optical frequencies \cite{maier2007plasmonics}. Compared with free-space radiation wavelength $\lambda_{0}$, the SP waves bounded to the interface can be characterized by a shorter effective wavelength $\lambda_{SP}$=$\lambda_{0}$/n$_{e}$ (where n$_{e}$ is the effective refractive index of the SP wave), giving rise to electromagnetic (EM) field localization and enhancement effects. This makes the nanostructures that can be used to couple resonantly an incident radiation with SP waves very attractive for a variety of applications \cite{berweger2012light,gramotnev2010plasmonics,liu2011nanoantenna,pan2011maskless,balvcytis2018fundamental,macdonald2010active,macdonald2009ultrafast,kristensen2017plasmonic,pelaz2017diverse,wang2017single}.

Utilization of cylindrical vector beams (CVBs) for excitation of plasmonic and all-dielectric nanostructures and their arrangements has recently become a subject of growing research interest \cite{sancho2012dark,yanai2014near,chen2012geometrical,gomez2013dark,zhan2009cylindrical}. In general, it is associated with the ability of pumping CVB efficiently to excite in such nanostructures converging SP waves \cite{gorodetski2008observation,kim2010synthesis,spektor2017revealing,chen2010experimental,tsai2014selective}, as well as specific resonant modes (dark plasmonic modes, Fano resonances, etc.) with larger Q-factors and local field enhancement. Such excitation modes cannot be easily reached with ordinary optical beams and are considered applicable   
for enhancing nonlinear optical interactions, sensing and catalytic performance, and so on. \cite{makarov2017light}.

Taking into account the spatially symmetric donut-shaped intensity profile of CVBs, resonant nanoantennas fabricated by them are commonly designed to have similar circular symmetry, for example, as circumferentially spaced nanodiscs \cite{yanai2014near,bao2015plasmonic} or slit arrangements and nanoapertures of various shapes \cite{chen2012geometrical,chen2012efficient,chen2009plasmonic,lerman2009demonstration}. However, experimental realization of such nanodevices typically requires utilization of rather expensive and time-consuming electron- and ion-beam lithography techniques. In this regard, complex-shaped plasmonic nanostructures designed and optimized to benefit from their excitation with spatially symmetric CVBs, and demonstrating the potential of being scaled up though easy-to-implement fabrication techniques, are of great demand for various applications.

Here, we demonstrate a simple, direct and maskless laser-based approach for fabrication of back-reflector-coupled plasmonic nanoring arrays. The approach is based on delicate ablation of an upper metal film of a metal-insulator-metal (MIM) sandwich with donut-shaped laser pulses followed by argon ion beam polishing. After being excited with a radially polarized beam, the as-fabricated MIM configuration of nanorings permits to realize efficient nanofocusing of constructively interfering plasmonic waves excited in the gap area between the nanoring and back reflector mirror. For the optimized geometric parameters of the produced MIM structure, substantial enhancement of electromagnetic near-fields at the center of the ring within a single focal spot with the size of 0.37$\lambda^{2}$ is shown to be achieved, which is confirmed by Finite Difference Time Domain (FDTD) calculations, as well as by detection of enhanced photoluminescence (PL) signal from adsorbed organic dye molecules. The simple, cost-efficient and easily scalable fabrication procedure, as well as its relatively good tolerance to excitation beam misalignment, make the newly-designed structures promising for realization of various nanophotonic and biosensing platforms based on CVB pump sources.

\textbf{METHODS.}

\begin{figure*}
  \includegraphics[width=0.85\textwidth]{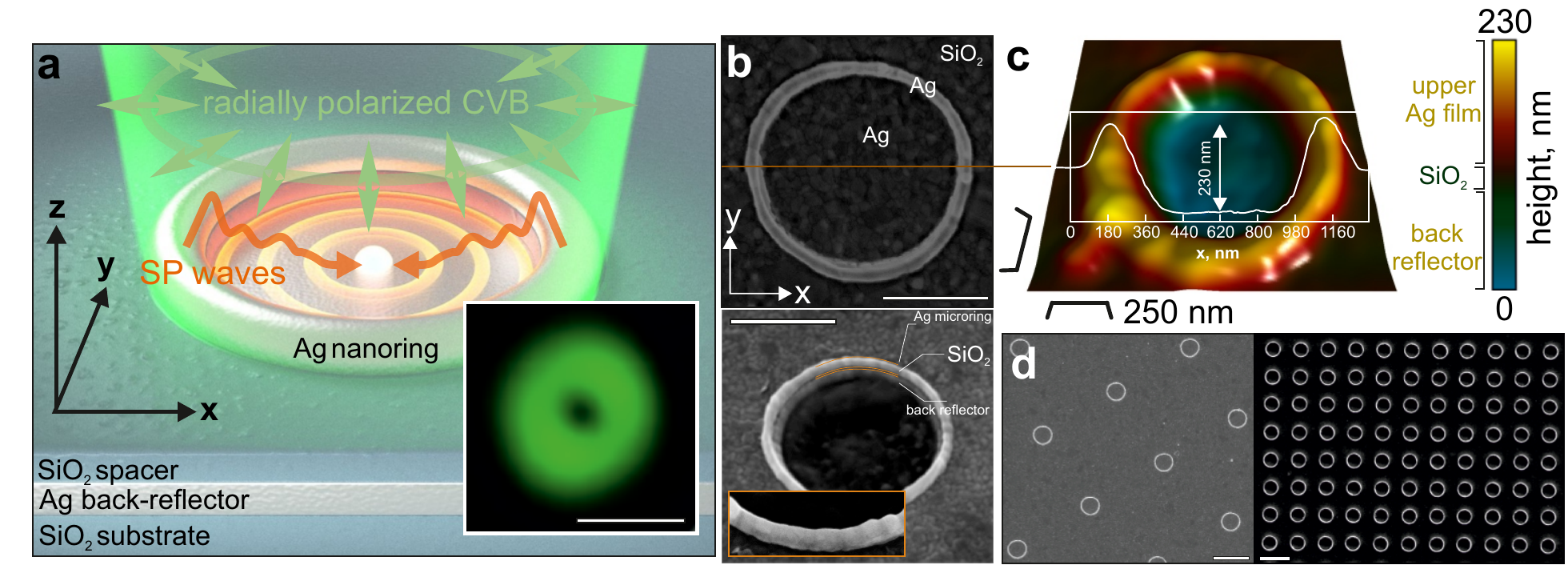}
  \caption{\textbf{Metal-insulator-metal plasmonic nanorings.} (a) Schematic presentation of MIM plasmonic nanoring irradiated with radially polarized CVB. Pump radiation excites SP waves in the gap area. Constructive interference of SPs produces localized and significantly enhanced electromagnetic hot spot in the geometric center of the structure. Inset shows measured focal-plane intensity pattern of a CVB beam used for fabrication of MIM nanorings and their excitation. (b) Representative top and side-view (view angle of 30$^o$) SEM images of MIM nanoring with diameter 1 $\mu$m, with scale bars indicating 500 nm. Inset in bottom image provides a close-up view on the nanoring's walls, showing their polycrystalline structure. (c) AFM image of a similar nanoring and its cross-sectional profile, showing its main vertical and lateral geometric parameters . (d) Top-view SEM image of several well ordered arrays of MIM nanorings printed at periods of 6 and 2 $\mu$m. Scale bar indicates 2 $\mu$m.}
  \label{fig1}
\end{figure*}

\textbf{Fabrication of MIM nanoring structures.}
First, a MIM sandwich was fabricated via consequent deposition of 400-nm thick Ag, 30-nm thick SiO$_2$ and 100-nm thick Ag films onto commercial Si wafer (Kurt Lescker) using high-vacuum e-beam evaporation. The sputtering rate was fixed at 50 nm$\cdot$min$^{-1}$ for Ag and 12 nm$\cdot$min$^{-1}$ for SiO$_2$, while sample holder was rotated at 25 rpm to ensure uniform film deposition. The thickness of each layer was controlled with a built-in microbalance system.

Then, the upper 100-nm thick Ag film was irradiated with second-harmonic (515 nm) 200-fs laser pulses generated by a regenerative amplified Yb:KGW-based laser system (Pharos, Light Conversion). The linearly polarized Gaussian-shaped intensity profile of the as-generated pulses were first converted to donut-shaped pulses using a commercial polarization converter (S-waveplate, Altechna, \cite{beresna2011radially}) and then focused onto the sample surface with a dry microscope objective having a numerical aperture (NA) of 0.8. The resulting focal plane intensity pattern with either a radial or azimuthal polarization of the generated donut beam is shown in Fig. 1a (inset), revealing its highly symmetric shape and a diameter of $\approx$0.95 $\mu$m. The produced microholes were then treated with a 7-mm diameter Ar ion beam (IM4000, Hitachi) at acceleration voltage of 3 kV, a discharge current of 105 $\mu$A and a gas flow rate of 0.1 cm$^3$/min. According to previous studies, such parameters provide melting-free slow-rate removal ($\approx$1 nm/s) of the Ag film \cite{kuchmizhak2014fabrication,kuchmizhak2016ion,syubaev2017fabrication}. This procedure allowed us to etch unmodified parts of the upper film, leaving isolated Ag nanorings on the surface of the SiO$_2$ spacer where the laser-modified Ag film was thicker. Importantly, the Ar$^+$ processing also removed the dielectric spacer inside the nanoring structure, while the beam shielding effect caused by the protruding Ag rim resulted in a concave surface profile of the crater inside the nanoring (see Fig. 1c).

The applied pulse energy $E$ of the shaped fs pulses was tailored with a motorized attenuator (Standa) and controlled with an energy meter (Nova, Ophir). The sample was set on a precise 3D positioning platform (Newport XM series) permitting to produce well-ordered arrays of nanostructures over the scanning area as large as 5x5 cm$^2$.

\textbf{Characterization of MIM nanorings.}
The morphology of the laser-ablated holes in the MIM substrate before and after their polishing with Ar$^+$ beam was carefully analyzed using high-resolution scanning electron microscopy (SEM, Ultra55+, Carl Zeiss) and atomic-force microscopy (AFM, NanoDST, Pacific Nanotechnologies).

\textbf{FDTD modeling.}
Numerical simulations of the MIM performance were carried out using a commercial 3D FDTD solver (Lumerical Solutions, Inc.). The import source modality was used to simulate either a radially or azimuthally polarized donut-shaped beam \cite{Note1} which irradiates the structure from the top at normal incidence. Donut beam diameter was varied between 0.9 and 1.2 $\mu$m to fit the size of the fabricated nanorings. Additionally, similar nanorings on a bulk glass substrate pumped with the CVB, as well as MIM structure with an ordinary linearly-polarized Gaussian-shaped beam, were also modeled for comparison. The main geometric parameters of the modeled MIM nanorings were extracted from the corresponding SEM/AFM analyses. For all calculations, the size of the elementary cubic cell was fixed at 1 nm$^3$, while the perfectly matched layers were utilized as boundary conditions limiting the computation volume. Dielectric functions of Ag and SiO$_2$ at 532-nm pump wavelength were extracted from reference \cite{Palik}. More details on the modeled geometry are given in Supporting Information Fig. S1.

\textbf{Plasmon-mediated PL enhancement measurements.} To study plasmonic nanofocusing and enhancement of the EM near-fields, the MIM nanorings were coated with a nm-thick layer of Rhodamine 6G (R6G) organic dye. For that, the sample was immersed into 10$^{-6}$ M R6G solution in ethanol for 2 h and then rinsed with deionized water. PL from R6G was excited with a donut-shaped radially (or azimuthally) polarized (532 nm) CVB, or with a Gaussian-shaped beam of the same diameter. The pump intensity was fixed at 10 $\mu$W/$\mu$m$^2$ to avoid degradation of emission at least within 2 min under laser irradiation \cite{kuchmizhak2016ion}. The diameter of the pump CVB in the focal plane of the 0.8-NA dry microscope objective was slightly adjusted by a lens system to fit the size of the MIM nanoring. The PL signal was collected with the same focusing system and analyzed with a home-built confocal system coupled to a grating-type spectrometer (Andor, Shamrock 303i) with a TE-cooled CCD-camera (Andor, Newton). Characteristic PL intensity distributions near the MIM nanorings were captured with a sensitive CCD camera. The sample was arranged on a piezo-positioning system to provide precise alignment of the donut beam with respect to the MIM nanoring.

\textbf{RESULTS AND DISCUSSION.}

\textbf{Fabrication of MIM nanorings.}
Schematic illustration of the laser-printed plasmonic nanofocusing element excited with a radially polarized CVB is presented in Fig. 1a. The proposed element is a micrometer-sized Ag ring separated from the underlying back-reflecting Ag mirror by a nm-thick SiO$_2$ spacer. To produce such a complex element, we adopted the previously developed laser-printing technology for fabrication of plasmonic nanorings \cite{kuchmizhak2016ion}.

The complex MIM geometry of the sample with a nm-thick dielectric spacer, which can be easily damaged or removed by laser pulses, imposes stringent requirements on the laser irradiation conditions. In particular, ordinary ns-long pulses with a Gaussian-shaped intensity profile used in previous studies were found to damage the dielectric spacer even at sub-threshold incident fluences. That is why, in this study we applied ultrashort fs laser pulses specifically shaped by S-waveplate polarization converter to have a donut-like intensity profile (see Methods section for details). The focal-plane intensity distribution of the beam used for ablation is shown in Fig.1a (inset). The combination of  ultrafast fs pulse excitation that aimed at minimizing heat affected zone and a more delicate impact on the irradiated Ag material in the central part of the donut-shaped focal spot was found to produce accurate through microholes in the upper film without any visible damage of the underlying SiO$_2$ spacer. The results of fs Gaussian-shaped printing of through holes at a fluence of 0.2 mJ/cm$^2$ are presented in Supporting Information Fig. S2, clearly illustrating the advantage of laser beam with donut-shaped intensity pattern for delicate ablation of layered samples.

The produced through holes were then processed with a 7-mm-diameter Ar-ion beam to provide melting-free slow-rate removal of the non-irradiated parts of the upper Ag film (see Methods for more details). The resulting surface morphologies representing isolated Ag microrings separated from the underlying Ag back-reflecting mirror by a SiO$_2$ spacer are shown in Fig. 1b. It should be noted that since both Ag and SiO$_2$ materials demonstrate comparative polishing rates, the processing Ar$^+$-beam also removed the dielectric spacer, as well as a part of the Ag back reflector inside the hole area, as it is revealed by the corresponding AFM image in Fig.1c.

\begin{figure*}
  \includegraphics[width=1\textwidth]{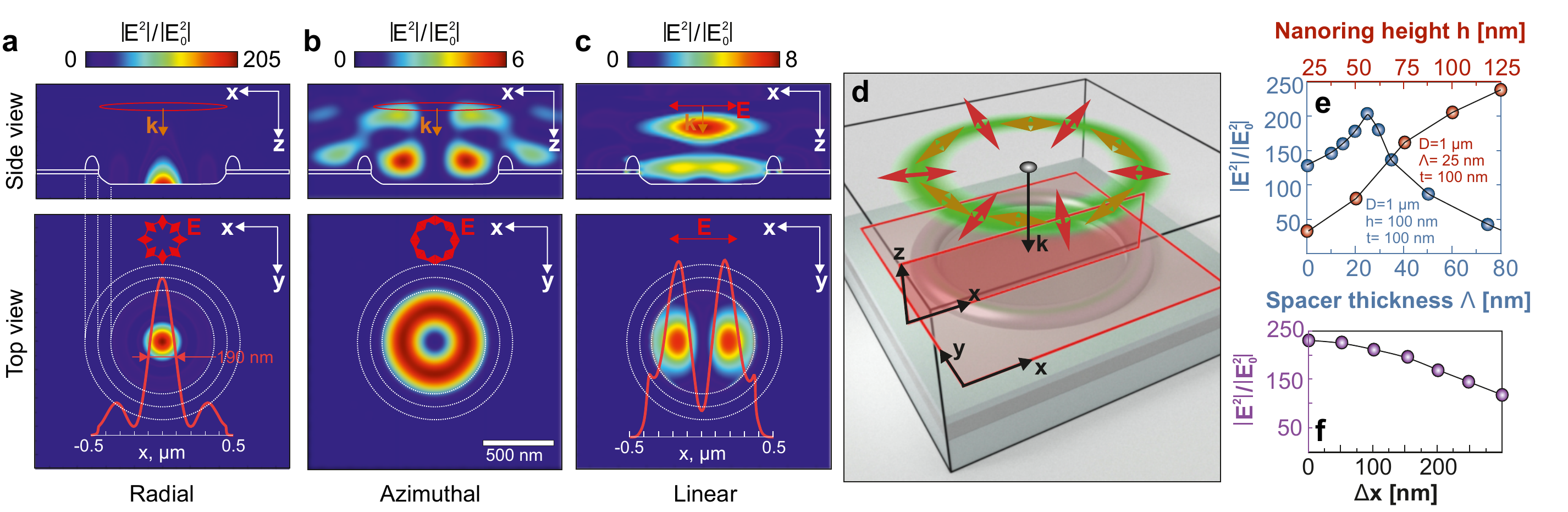}
  \caption{\textbf{FDTD modeling of MIM nanorings.} Normalized squared electric field amplitude $\mid$E$^2\mid$/$\mid$E$_0^2\mid$ calculated near MIM nanoring pumped with (a) radially and (b) azimuthally polarized CVB with diameter 1 $\mu$m and excitation wavelength 532 nm. (c) Similar calculations performed for  MIM nanoring pumped with linearly (x) polarized Gaussian beam of 1 $\mu$m in diameter. Geometrical parameters of the modeled MIM nanoring are $D$= 1 $\mu$m, $h$=100 nm, $t$=100 nm and $\Lambda$=25 nm. White curves highlight the geometry of the nanoring. Top-view images are calculated in the plane situated 5 nm above the crater inside the nanostructure. Polarization direction in each case is schematically illustrated by red arrows. The color bar limits differ for each case. (d) Schematic of the simulated geometry. The red-color rectangles indicate the position of the EM monitors. (e) Maximal normalized squared electric field amplitude $\mid$E$^2\mid$/$\mid$E$_0^2\mid$ in the plasmonic focal spot as a function of SiO$_2$ spacer thickness $\Lambda$ (blue markers) and the nanoring height $h$ (red markers). (f) Maximal normalized squared electric field amplitude $\mid$E$^2\mid$/$\mid$E$_0^2\mid$ versus the CVB lateral misalignment $\Delta$x with respect to the nanoring center.}
  \label{fig1}
\end{figure*}

As expected, the main geometrical parameters of the produced nanorings (i.e., diameter $D$, height $h$, wall thickness $t$, and spacer thickness $\Lambda$, see Supporting Information Fig. S1) were found to depend on both the focusing/energetic characteristics of the shaped laser pulses and on the thickness $d$ of the upper Ag film. In this paper, we do not provide detailed studies of these dependencies which are to be the topic of our forthcoming studies. However, we will further discuss the general trends and main features related to utilization of shaped fs pulses, which differs the present work from previously published results.

First, irradiation of a noble metal film with $d$ thicker than 150 nm with a donut-shaped fs pulse at subthreshold fluences is known to cause the thermomechanical spallation of the partially molten upper layer of the film \cite{kuchmizhak2016fly,pavlov2018ultrafast}. Such spallation provides a rough spiky surface which is definitely out of interest for precise fabrication of nanorings targeted in this study. In this respect, we utilized the upper Ag film with $d$=100 nm to avoid spallation effect. According to our previous studies, the initial thickness of the irradiated film governs the maximal nanoring height above the unmodified Ag film level \cite{kuchmizhak2016ion}. For the given laser printing conditions and $d$=100 nm, the average maximal height $h$ for fs-pulse-printed microrings was found to be about 130 nm, according to AFM studies (Fig.1c). In addition, the resulting height $h$ of the laser-printed nanorings can be further reduced by tailoring the duration of the Ar$^+$-beam treatment.

The nanoring diameter $D$ generally can be tuned by changing the donut-shaped beam size, as well as the pulse fluence. However, taking into account the above discussed MIM configuration of the structure, only rather narrow fluence range slightly above the ablation threshold (0.3$<F<$0.36 mJ/cm$^2$) can be applied to process a 100-nm thick upper Ag film to avoid damaging the underlying dielectric spacer. In this respect, the minimal size of the produced nanorings is limited by the focal-plane size of the donut-shaped beam (which is $\approx$ twice larger compared with the size of the Gaussian spot). For our dry microscope objective with NA=0.8 and the above mentioned fluence range, we managed to produce nanorings with diameters $D$ between 0.9 and 1.2 $\mu$m.

For the chosen experimental conditions, the thickness of the produced nanorings $t$ was found to be about 100 nm, which is comparable with that of nanorings produced via ns-pulse printing. However, the use of fs-long pulses was shown to minimize both the thickness $\Delta t$ and height variation $\Delta h$ of the nanoring walls. Systematic AFM and SEM measurements demonstrated the average values of $\Delta h$ and $\Delta t$ for the fs-laser printed microrings to be around 20 nm and 30 nm, respectively.

The good pulse-to-pulse stability (better than 0.5\%) of the commercial fs-pulse laser system used and its sub-MHz pulse repetition rate ensured ultrafast reproducible fabrication of well ordered arrays of nanorings (see Fig. 1d). Finally, it should be mentioned that the above described approach is quite flexible toward materials for designing and fabricating such MIM structures. The upper microrings can be potentially produced using all common plasmonic films, such as gold, silver and copper, as well as their multi-component alloys, while various transparent and reflecting materials can be used as spacer and back reflector, respectively. The thickness of the spacing layer can be easily controlled during its deposition by means of e-beam. However, in this paper we only used SiO$_2$ spacer layer with its thickness $\Lambda$ of 25 nm as this provided the best performance according to our modeling results presented further.

\textbf{FDTD modeling of the nanoring characteristics.}
 The performance of the above described MIM nanorings with variable geometry was first characterized via comprehensive FDTD modeling of electromagnetic near-fields excited by either a radially or azimuthally polarized CVB (see Methods for details). For each set of geometric parameters of the modeled structure and excitation conditions, the obtained E-field maps were used to access the maximal enhancement of the normalized squared electric field amplitude $\mid$E$^2\mid$/$\mid$E$_0^2\mid$. For comparison, we also performed similar computations for isolated nanorings with the same geometry and size placed on semi-infinite glass substrate. The results of these calculations are shown in Supporting Information Fig.S3.

Figure 2 provides representative 2D maps showing distribution of $\mid$E$^2\mid$/$\mid$E$_0^2\mid$ near a 1-$\mu$m-sized MIM nanoring pumped with radially and azimuthally polarized CVBs, as well as with a linearly-polarized Gaussian beam of the same diameter and excitation wavelength 532 nm. The calculated maps clearly show remarkable enhancement of the $\mid$E$^2\mid$/$\mid$E$_0^2\mid$ value within the central spot in case of radially polarized CVB pump mediated by focusing of constructively interfering SP waves (see Fig.2a). Such interference produces a rather confined spot having 1/e$^2$-radius R$_{SP}$ of 140 nm resulting in the focal size of $\pi$R$_{SP}^2$=0.062 $\mu$m$^2$ (0.37$\lambda_0^2$), which is substantially smaller when compared with the focal spot size $\approx\pi$(0.61$\lambda_0$/NA)$^2$=0.364 $\mu$m$^2$ (1.3$\lambda_0^2$) provided by a dry lens at NA=0.95 and $\lambda$=532 nm. In the contrary, neither azimuthally polarized CVB nor linearly polarized Gaussian beam pump were found to be capable of achieving substantial field enhancement and plasmonic focusing (Fig.2b,c). Decomposition of the calculated field in Fig. 2a into orthogonal field components demonstrates that over 90\% of the total E-field amplitude corresponds to the E$_Z$-field component directed normally with respect to the sample surface. In particular, this confirms that focusing effect is caused by excitation and constructive interference of SP waves.

The maximal normalized squared electric field amplitude $\mid$E$^2\mid$/$\mid$E$_0^2\mid$ in the plasmonic focal spot was further calculated as a function of SiO$_2$ spacer thickness $\Lambda$ (blue curve, Fig. 2e). The calculations gave the optimal spacer thickness of $\Lambda$= 25 nm for the fixed geometric parameters of the nanoring (D= 1 $\mu$m, $h$= 100 nm and $t$= 120 nm). As a result, the theoretical maximal value of $\mid$E$^2\mid$/$\mid$E$_0^2\mid$ was found to be $\approx$  200. In a similar way, for a fixed optimal thickness of dielectric spacer ($\Lambda$=25 nm), we also modeled the $\mid$E$^2\mid$/$\mid$E$_0^2\mid$ versus the height $h$ of the nanoring walls (red dots, Fig. 2d). According to our experimental data, the maximal height of produced nanorings measured above the initial level of the laser-treated 100-nm thick Ag film reached 120$\pm$20 nm, and could be further reduced in the process of Ar$^+$-beam polishing. The performed simulations showed the maximal theoretical value of $\mid$E$^2\mid$/$\mid$E$_0^2\mid$ could achieve as much as 240 at $h$=125 nm.

\begin{figure*}
  \includegraphics[width=0.85\textwidth]{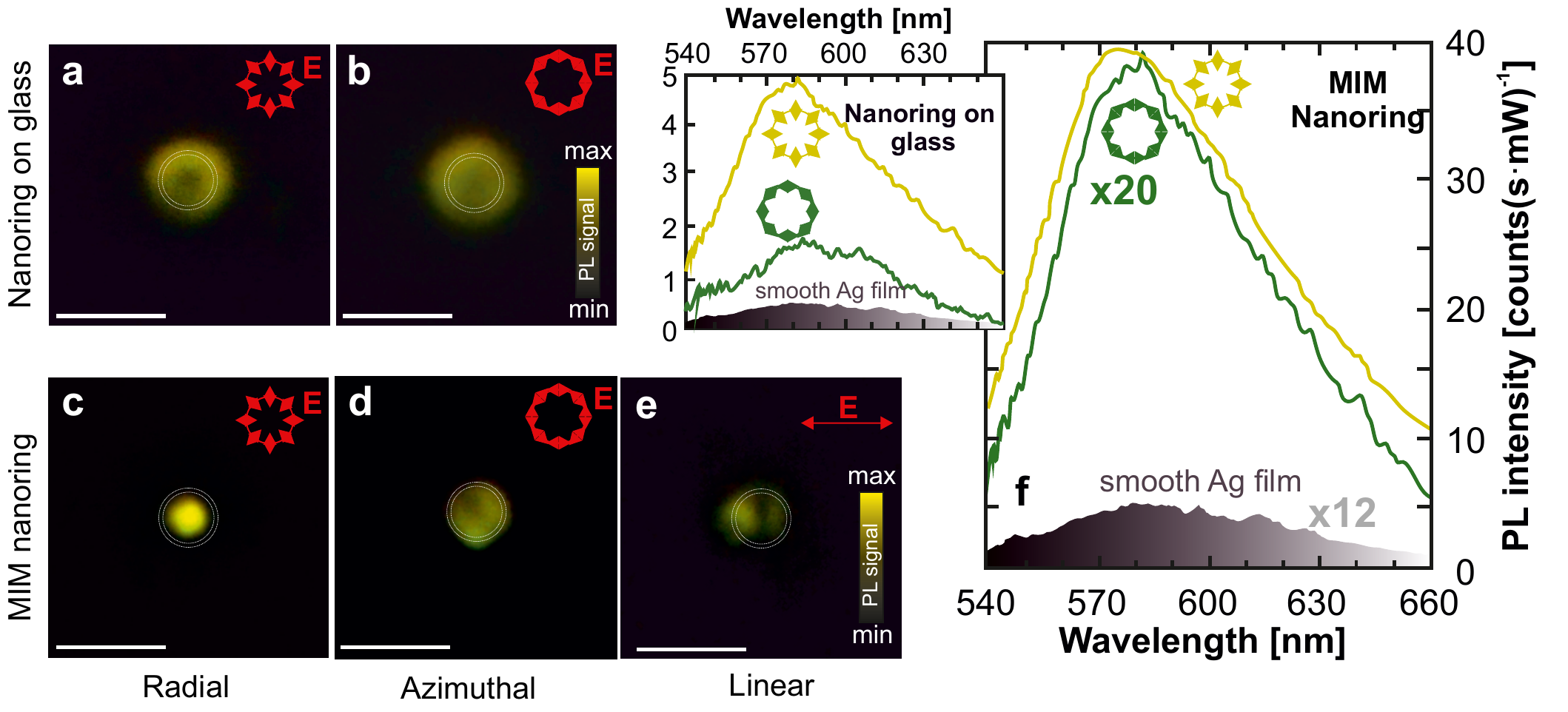}
  \caption{\textbf{Plasmonic performance of the MIM nanorings.} (a,b) R6G PL images of the isolated 1-$\mu$m diameter Ag nanoring on a glass substrate irradiated with radially and azimuthally polarized CVB at pump intensity I=10 $\mu$W/$\mu$m$^2$. (c-e) Similar PL images obtained by pumping the MIM nanoring structure of same size with radially and azimuthally polarized CVB as well as with a linearly polarized Gaussian beam. Red arrows indicate the polarization direction. Intensity of each PL image is homogenized by changing the accumulation time for better displaying. Nominal geometric dimensions of the tested nanorings are as follows: $D$= 1 $\mu$m, $h$=100$\pm$20 nm, $t$=100$\pm$20 nm, $\Lambda$=25 nm. Scale bar corresponds to 2 $\mu$m. (f) R6G PL spectra measured from the MIM nanoring excited with a radially and azimuthally polarized CVB. Inset shows similar measurements obtained for isolated Ag nanoring on a glass substrate. Different scaling factors of the spectra are used for better displaying. The R6G PL signal obtained from the smooth nanocrystalline Ag film by linearly polarized Gaussian beam is provided as a reference.}
  \label{fig3}
\end{figure*}

It is worth noting that within the range of modeled parameters (mostly defined by the geometry of  fabricated MIM nanorings), the size of plasmonic focal spot $\pi$R$_{SP}^2$ remained constant irrespective of the variation of main geometric parameters of the nanorings (spacer thickness $d$, diameter $D$, as well as the wall thickness $t$ and height $h$). However, as the performance of the proposed structure relies on the interference of SP waves, the maximal EM-field enhancement in the plasmonic ``hot spot'' is expected to vary substantially with the nanoring diameter $D$ and scale with the SP wavelength $\lambda_{SP}$.  According to our experimental data, the range of available diameters $D$ of the MIM nanorings lay between 900 and 1200 nm. Our simulations show that for the chosen MIM geometry, the maximal enhancement should be achieved at $D\approx$1 $\mu$m, while any variation of $D$ within the available range resulted in a decrease of the $\mid$E$^2\mid$/$\mid$E$_0^2\mid$ in the central focal spot. Details of this modeling are provided in Supporting Information Fig.S1.

Finally, the performance of the fabricated nanorings was found to show tolerance to slight misalignment $\Delta$(x) of the pumping CVBs. To illustrate this useful feature, we calculated the maximal value of $\mid$E$^2\mid$/$\mid$E$_0^2\mid$ and the focal spot size for optimized MIM nanoring pumped with a radially polarized CVB that gradually shifted along the x-axis with respect to the nanoring center. As clearly seen in Fig.2f, a substantial CVB misalignment $\Delta$x equal to 250 nm provided about twice lower EM-field enhancement, which is comparable with the case of aligned pumping, while the shape and size of the plasmonic focal spot remained unchanged.

\textbf{Enhancement of R6G spontaneous emission on CVB-pumped MIM nanorings.}

To verify experimentally the formation of a bright and spatially confined plasmonic ``hot spot'' upon excitation of the nanorings with a CVB, we coated a 10-nm thick layer of R6G organic dye above the fabricated structure (see Methods for more details). The MIM nanorings were fabricated to have their geometrical parameters optimized via the above mentioned numerical simulations. Additionally, the isolated nanorings on a glass substrate are also used to support the obtained results. Spontaneous emission from the adsorbed molecules pumped with either a radially or azimuthally polarized CVB is collected with the same high-NA dry microscope lens. Such experimental scheme allows to simultaneously visualize the characteristic position of the plasmonic ``hot spots'' near the pumped nanorings at a diffraction-limited lateral resolution of $\approx$ 300 nm using a CCD-camera as well as to quantitatively analyzed the overall R6G PL signal from the area of single nanostructure by a grating-type spectrometer.

Figure 3a,b shows PL images obtained from the R6G layer deposited onto an isolated nanoring (fabricated on glass substrate) and excitated with radially and azimuthally polarized CVBs. As clearly seen, the characteristic PL intensity distribution near the nanoring generally replicated the donut-shaped CVB and agreed with the FDTD simulation results that predicted EM-field enhancement only near the nanoring walls for both polarization states (see Fig. S3 in Supporting Information). However, for radially polarized CVB the R6G PL yield appears to be enhanced by a factor of 2.5 when compared with the case of azimuthally polarized excitation, which is also well consistent with the numerical simulations.

In a sharp contrast to the case of isolated nanorings, their MIM arrangement is seen in Fig.3c-f to demonstrate a substantial difference in both its characteristic PL distribution and its averaged PL yield from the capping R6G layer when the polarization of pumping beam was varied. More specifically, for the constant pump intensity of I$_{pump}$=10 $\mu$W/$\mu$m$^2$, about 20-fold and 100-fold enhancement of PL signal was registered, respectively, when radially- and azimuthally- polarized CVBs were applied as excitation source. The obtained spectra are also supported by representative R6G PL images of the isolated MIM nanoring pumped with radially and azimuthally polarized CVBs, as well as with a linearly polarized Gaussian beam. As well seen in Fig.3, for radially polarized CVB pumping, spontaneous emission from R6G was predominantly excited in the central plasmonic "hot spot'', while some contribution from molecules adsorbed on the nanoring walls could also be expected.

\textbf{CONCLUSIONS AND OUTLOOK.}
In conclusion, we proposed a facile maskless laser-based technology that permits to fabricate plasmonic nanorings separated from the underlying back-reflecting mirror by a thin dielectric spacer, i.e. MIM nanorings. The technology is based on delicate ablation of the upper metal film of a metal-insulator-metal sandwich with donut-shaped laser pulses followed by polishing with an Ar$^+$ beam. The ability of the fabricated MIM nanorings to produce spatially confined plasmonic "hot spots'' upon excitation with radially polarized CVBs is revealed via comprehensive FDTD modeling and validated by detecting spontaneous emission from R6G organic dye placed inside the nanorings. When properly pumped, an isolated MIM nanoring is shown to provide $\approx$100-fold enhancement of PL emission from R6G when compared with the smooth nanocrystalline Ag metal film. It is believed that the simple, reproducible and easily scalable procedure proposed in this study will find further applications in the future as a very efficient approach to produce a wide number of devices for  nanophotonic applications.
As a potential follow-up of this work, one can suggest a plasmonic nanoantenna placed directly into the plasmonic ``hot spot'' generated by the above described MIM nanoring (see Supporting Information Fig. S5). Our preliminary calculations showed the EM field enhancement as large as 10$^6$ can be achieved underneath 100-nm-sized spherical Ag nanoparticles used as such nanoantennas that could be potentially transferred inside the plasmonic focal spot of the MIM nanoring by means of either scanning probe microscopy or laser-induced forward transfer techniques \cite{kuznetsov2011laser,unger2012time,visser2015toward}.

{\begin{acknowledgments}
Authors acknowledge the support of the Russian Science Foundation (grant no. 17-12-01258).

\end{acknowledgments}}
\end{document}